\documentclass[twocolumn,showpacs,aps,prd,preprintnumbers,amsmath,amssymb,superscriptaddress,groupedaddress,nofootinbib]{revtex4}
\usepackage{graphicx,subfigure,multirow}
\usepackage{dcolumn,lipsum}% Align table columns on decimal point
\usepackage{bm}% bold math
\usepackage{natbib}
\usepackage{doi}%<----------
\usepackage{soul}%this package is for text strike out
\usepackage[utf8]{inputenc}
\makeatletter

\newcommand{\fmslash}[2][0mu]{%
  \mathchoice
    {\fmsl@sh\displaystyle{#1}{#2}}%
    {\fmsl@sh\textstyle{#1}{#2}}%
    {\fmsl@sh\scriptstyle{#1}{#2}}%
    {\fmsl@sh\scriptscriptstyle{#1}{#2}}}
\newcommand{\fmsl@sh}[3]{%
  \m@th\ooalign{$\hfil#1\mkern#2/\hfil$\crcr$#1#3$}}
\makeatother

\newcommand{\lsim}{{\;\raise0.3ex\hbox{$<$\kern-0.75em\raise-1.1ex\hbox{$\sim$}}\;}}
\newcommand{\gsim}{{\;\raise0.3ex\hbox{$>$\kern-0.75em\raise-1.1ex\hbox{$\sim$}}\;}}

\newcommand{\met}{{\fmslash E_T}}
\usepackage{color}
\usepackage{url}

\begin{document}
\title{Demystifying the compressed top squark region with kinematic variables}
\author{Partha Konar}
\email{konar@prl.res.in}
\affiliation{Physical Research Laboratory (PRL), Ahmedabad - 380009, Gujarat, India}
\author{Tanmoy Mondal}
\email{tanmoymondal@hri.res.in}
\affiliation{Physical Research Laboratory (PRL), Ahmedabad - 380009, Gujarat, India}
\affiliation{Regional Centre for Accelerator-based Particle Physics, Harish-Chandra Research Institute,
  HBNI, Chhatnag Road, Jhusi, Allahabad - 211019, India}
\author{Abhaya Kumar Swain}
\email{abhaya@prl.res.in}
\affiliation{Physical Research Laboratory (PRL), Ahmedabad - 380009, Gujarat, India}
\date{\today}

% = = = = = = = = = = = = = = = = = = = = = = = = = = = = = = = = = = = = = = = = = = = = = = = = = 
\begin{abstract}
The ongoing perplexing scenario with no hints of new physics at the Large Hadron Collider can be elucidated amicably if the exotic particle spectrum in many of the well-motivated theoretical models possesses degenerate mass. We investigate the usefulness of different kinematic variables sensitive to the compressed mass region, and propose a search strategy considering a phenomenological supersymmetric scenario where the top squark undergoes a four-body decay due to its extremely narrow mass difference with the lightest supersymmetric particle.  
%Considering a challenging but relatively clean decay channel,we demonstrate that one can effectively restrain the significant background from top  {\color{blue}providing a complementary approach to the present CMS analysis that can be used to extend the current limit to a phase-space region not yet explored. }
Considering a challenging but relatively clean dileptonic decay channel, we demonstrate that one can effectively restrain the significant background from the top quark, which provides a complementary approach to the present CMS analysis. With the new strategic approach the current limit can be extended to a phase-space region that was not explored before.

\end{abstract}

% 14.80.Ly Supersymmetric partners of known particles, 12.60.Jv Supersymmetric models  
%13.85.?t Hadron-induced high- and super-high-energy interactions

\maketitle

% = = = = = = = = = = = = = = = = = = = = = = = = = = = = = = = = = = = = = = = = = = = = = = = = = 
%%%%%%%%%%%%%%%%%%%%%%%%%%%%%%%% 
\section{Introduction}
\label{sec:intro}
%%%%%%%%%%%%%%%%%%%%%%%%%%%%%%%%
%{\textit{Introduction:-} 
The Large Hadron Collider (LHC) with its enhanced center of mass energy and the luminosity holds phenomenal potential to search for physics beyond the Standard Model (BSM). Among possible extensions, the supersymmetry (SUSY) is undoubtedly the most appealing theory waiting to be discovered at the LHC. It  naturally stabilizes the Higgs boson mass against large quantum correction with light top squark mass ($\le 1$ TeV). Searching for the natural SUSY at the LHC is challenging since the final states involve at least two invisible massive lightest supersymmetric particles (LSPs) that escape detection. These LSPs are popularly considered as potential dark matter candidate in an R-parity conserved model. There are many dedicated prescriptions discussed in the literature using which the LHC severely constrained the light top squark mass ($m_{\tilde{t}}$). The direct searches exclude $m_{\tilde{t}}$ below $800-900$ GeV when the top squark ($\tilde{t}$) decays to a top quark and a neutralino~\cite{ATLAS-CONF-2016-050,CMS-PAS-SUS-16-028}.

The SUSY in a multi-TeV domain, albeit at the price of naturalness, can further be probed in very high energy collider. However, a more prudent approach might be required to scrutinize all such hitherto unexplored possibilities, if the top squark could still be hidden inside the current collider data. A particular phenomenological choice on masses, lacking any knowledge of an actual SUSY breaking mechanism, can provide such a scenario where the LHC exclusion bounds are particularly poor. This region of the SUSY spectrum  is popularly known as  the  compressed region,\footnote{One can also consider full supersymmetry with a sufficiently compressed spectrum satisfying all available constraints, such as from Higgs measurement and dark matter~\cite{Dutta:2015exw, Dutta:2017jpe}.} where the mass difference between the $\tilde{t}$ and the LSP, commonly taken as the neutralino ($\chi$), is small. The small mass gap leads to the production of soft particles making it very difficult to identify them in the detector. Moreover, the massive neutralinos carry the highest fraction of the top squark momentum and each of them flies in the opposite direction, leading to cancellation of the transverse momentum between them. Consequently, the characteristic SUSY signature of large  missing transverse momentum ($\met$)  is not potent enough to size the background events. In order to detect  soft particles from the signal region and also to produce a sizable amount of missing transverse momenta, one is required to have reasonably high $P_T$ initial state radiation (ISR) jet(s) accompanying the top squark pair production. 

Depending upon the smallness in $\tilde{t}-\chi$ mass gap, different decay channels and thus scope for various 
search schemes for the $\tilde{t}$ arise.  For example, if $\Delta M \equiv (m_{\tilde{t}}-m_\chi) < m_W+m_b$, 
the top squark can decay via the flavor changing neutral current through loop-induced two-body decay mode
$\tilde{t}\to\,c\,\chi$ or the four-body mode
$\tilde{t}\to\,b\,f\,f'\,\chi$~\cite{Hikasa:1987db, Boehm:1999tr, Muhlleitner:2011ww}.

In the two-body decay, since the charm  quark cannot be tagged efficiently inside a jet, both the CMS~\cite{Khachatryan:2015wza,Khachatryan:2016dvc} and the ATLAS~\cite{Aad:2014nra,Aad:2015pfx} rely on the monojet+$\met$ signal  with 8 TeV energy.  With new 13 TeV data CMS  used the $\alpha_T$ variable and the limit on top squark mass goes up to 400 GeV with neutralino mass of 310 GeV. A conventional monojet+$\met$ search with 13 TeV data provide limit on top squark mass of 323 GeV as reported by the ATLAS collaboration~\cite{Aaboud:2016tnv}. 

For the leptonic four-body top squark decay $\tilde{t}\to\,b\,\ell\,\nu_\ell\,\chi$, LHC collaborations have carried out  different searches.  In these searches at least one lepton is reconstructed where the full signal consists of lepton(s)+jets+$\met$.  With the 8 TeV data both the CMS~\cite{Khachatryan:2015pot} and the ATLAS~\cite{Aad:2014kra} have searched the compressed region with only one lepton in the final state. The CMS moved one step further with the new 13 TeV data  and explored the region with both one lepton~\cite{CMS-PAS-SUS-16-031} and two leptons~\cite{CMS-PAS-SUS-16-025} in the final state. Assuming a 100\% branching ratio of the four-body  decay, and prompt decay top squark masses below 330 GeV are excluded at 95\% confidence level for a mass difference to the LSP of about 25 GeV in one lepton search~\cite{CMS-PAS-SUS-16-031}, whereas the limit from  the dilepton search is 360 GeV with $\Delta M = 30$ GeV~\cite{CMS-PAS-SUS-16-025}. However, the mass limit weakens drastically with larger mass gap and  for $\Delta M \simeq m_W$ the limit slips down to 270 GeV~\cite{CMS-PAS-SUS-16-025}. In this work we demonstrate that the judicial use of kinematic variables can improve these limits, in particular, for larger mass gap in the four-body region.

Several novel methods were proposed to search for the compressed region at the collider~\cite{Chou:1999zb,Das:2001kd,Carena:2008mj,Bornhauser:2010mw,Ajaib:2011hs,Kats:2011it,He:2011tp,Drees:2012dd,Belanger:2012mk,Alves:2012ft,Han:2012fw,Choudhury:2012kn,Dreiner:2012sh,Krizka:2012ah,Delgado:2012eu,Hagiwara:2013tva,Belanger:2013oka,Dutta:2013gga,Czakon:2014fka,Grober:2014aha,Ferretti:2015dea,Rolbiecki:2015lsa,Macaluso:2015wja,Kaufman:2015nda,Kobakhidze:2015scd,Goncalves:2016nil,Goncalves:2016tft,Hikasa:2015lma,Duan:2016vpp,Cho:2014yma}. Recently, an interesting but simple kinematic variable $R_M$~\cite{An:2015uwa} was proposed that suited the compressed region of SUSY.  It needs a hard ISR jet to be produced with a top squark pair and is defined as the ratio between missing transverse momenta and the ISR jet transverse momenta. This variable peaks at the neutralino and top squark mass ratio (${m_{\chi}} / {m_{\tilde{t}}}$) while the background falls exponentially. Subsequently, it is noticed that with the presence other sources of missing energy that may come from the neutrino(s) of leptonic decay modes, this variable spreads around the peak leading to the reduction of the signal and background discriminating power. For the semileptonic decay, the neutrino contribution can be subtracted~\cite{Cheng:2016mcw}, which can restore the behavior of the $R_M$ variable but for the dileptonic decay channel separation of neutrino contribution is not possible.

We focus mainly on  the $\Delta M < (m_W+m_b)$ mass gap region and exploit the suitable kinematic variables to constrain the parameter space still untouched by the  ATLAS and the CMS collaborations. The decay products from the top squark include both visible and invisible particles and any constructed observable, lacking the full phase-space information, exhibits kinematic singularities in the observable phase space~\cite{Kim:2009si}. These observables exhibit strikingly different behavior for the signal and the background events and possess the potential to provide/extend the limit on the top squark mass for the compressed scenario.

In this article we have shown that for the leptonic decay channel, our proposed kinematic variables can complement the state-of-the-art limits on the top squark mass provided by the CMS collaboration. These variables are capable of extending the CMS limit of 270 GeV to 335 GeV with existing  13 $fb^{-1}$ data. Also, we emphasize that the limits obtained in the leptonic channel are comparable to the results obtained in the full hadronic channel with the variable $R_M$.

% = = = = = = = = = = = = = = = = = = = = = = = = = = = = = = = = = = = = = = 
% = = = = = = = = = = = = = = = = = = = = = = = = = = = = = = = = = = = = = = 
% = = = = = = = = = = = = = = = = = = = = = = = = = = = = = = = = = = = = = = 
%%%%%%%%%%%%%%%%%%%%%%%%%%%%%%%% 
\section{Kinematic and invariant mass variables}
\label{sec:variables}
%%%%%%%%%%%%%%%%%%%%%%%%%%%%%%%%
%\textit{Kinematic and invariant mass variables:- } 
In order to demonstrate the efficacy of kinematic features,
we consider a challenging but clean dileptonic channel,
\begin{equation}
PP \rightarrow \tilde{t} + \tilde{t}^* \rightarrow \chi_1^0\, b\, \ell^{+}\, \bar{\nu_{\ell}} + \chi_1^0\, \bar{b}\, \ell^{-}\, \nu_{\ell},
\end{equation}
along with ISR jet(s). Evidently, the signal we consider for our analysis contains two leptons, at least one b-tagged jet, one or more high $P_T$ ISR jet(s), and large missing transverse momenta. Since all the leptons and $b$ quarks are mostly soft, it is not very economical to tag both the $b$ jets due to low b-tagging efficiency. The leading background for this signal region is the top pair production whereas subleading contribution comes from associated $tW$ production.  Since we are tagging only one $b$ jet, other possible background can come from semileptonic decay of a top and the other nonprompt lepton from the $B$-meson decay. We find that the contribution is negligible($< 5\%$) compared to our leading backgrounds.  
   
%%====================================
\begin{figure}[t] 
\includegraphics[width=7.5cm]{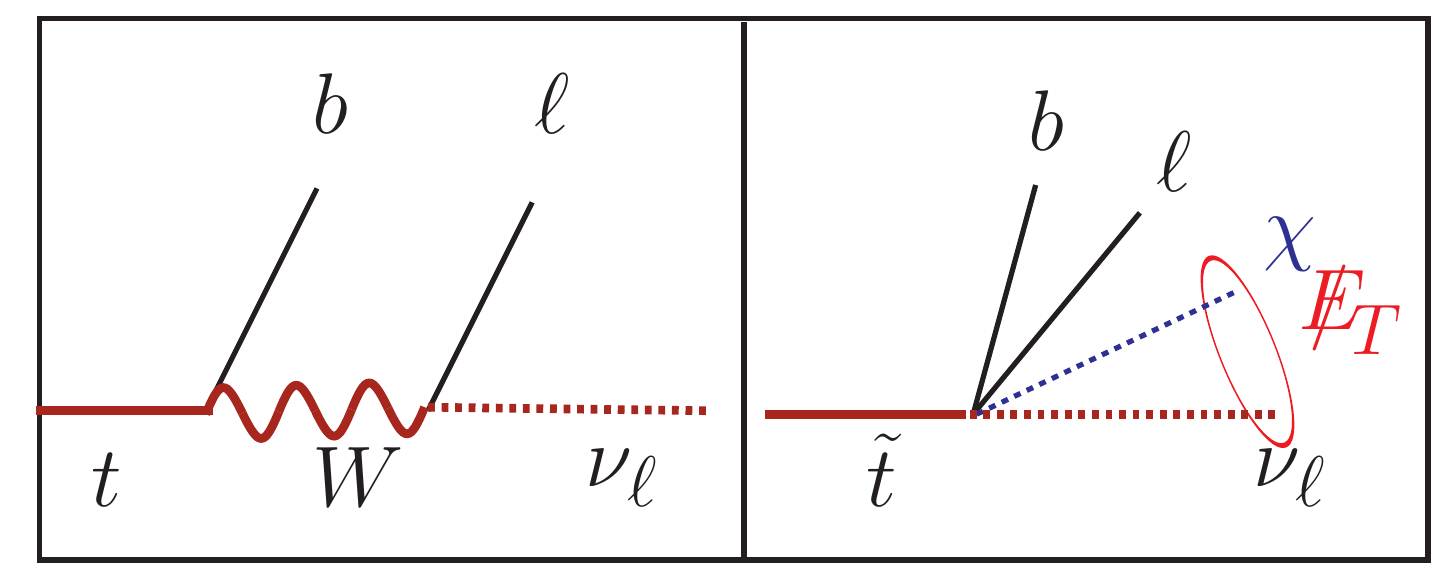}
\caption{The right panel shows the top squark four-body decay producing top and $W$boson off shell. The left panel shows the top decay via the leptonic channel.
}
\label{fig:SigAndBackgTopology}
\end{figure}
%%====================================

The cascade decay topology for both the signal and background is shown in the Fig.~\ref{fig:SigAndBackgTopology}. In this scenario both the top and subsequently the $W$ boson are produced off shell for the signal resulting in the four-body decay, unlike the background events. In the signal region, the two invisible particles, from each top squark decay, are combined to form an effective invisible particle as represented by the oval in Fig.~\ref{fig:SigAndBackgTopology} with invariant mass $m_{I}$. 

These distinct kinematic topologies between the signal and the background empower one to look for different kinematic variables possessing characteristic observable singularities in phase space to discover or exclude the light top squark in the compressed region at the LHC. The kinematic variables that best incorporate the topology information are the ones having the best discriminating power, {\it e.g.},   the visible invariant masses~\cite{Cho:2012er}. In our present example invariant mass of the b-tagged jet and the lepton, $M_{b \ell}$, can be utilized for maximizing the signal to background ratio. The distribution of the variable $M_{b \ell}$ has an end point that arises, as mentioned earlier, because of the singularity in the observable phase space. The full phase space does not include a singularity but observable phase space does as we measure a subset of event momenta~\cite{Kim:2009si}. The invariant mass $M_{b \ell}$ is a projection of full phase space on to the observable phase space and any folding in the full phase space resulted in a singularity (end point). Position of the edge of the $M_{b \ell}$ distribution depends on the decay topology as
\begin{equation}
M_{b \ell}^{max} = \left \{ 
   \begin{aligned}
   & \sqrt{m_t^2 - m_W^2},~~\text{for background}\\
   & (m_{\tilde{t}} - m_I^{min}) = \Delta M ,~~\text{for signal,}
   \end{aligned}
   \right.
\end{equation}
after neglecting the neutrino mass. Therefore, for the background events, the position of the end point of $M_{b \ell}$ distribution is larger compared to the signal events that are confined within $\Delta M$.

At this point, we also propose two new ratios possessing a distinct facet specifically for the four-body decay of the compressed region of SUSY. To motivate with concrete examples, one starts with a scenario with mass difference between top squark and neutralino being tiny, such as, $\Delta M = 5~\text{GeV}$. The transverse momenta of the top squark and $b$ jet are related as $P_{T}^{\tilde{t}} = ({m_{\tilde{t}}}/{m_b}) P_T^b$. One can also write similar equations for the corresponding lepton from the decay and finally, using both these relations we construct two new ratios, 
\begin{eqnarray}
R_{b E} = \frac{\sum P_T^{b_i}}{\met}, \; \; \; \; \; 
R_{\ell E} = \frac{\sum P_T^{{\ell}_i}}{\met}.
\end{eqnarray}
It is easy to follow that for the signal region $R_{bE}$ peaks at the mass ratio $({m_b}/{m_{\chi}})$, whereas $R_{\ell E}$ peaks at $({m_{\ell}}/{m_{\chi}}) \approx 0$. We show that these two interesting ratios are better suited for the dileptonic decay channel exploring the top squark four-body decay scenario.

Our final observable is the stransverse mass $M_{T2}(b \ell \ell)$~\cite{Lester:1999tx,Konar:2009qr,Burns:2008va,Barr:2011xt}, which was also favored with good discriminating power.
Although the symmetric $M_{T2}$ constructed from  two $b$ jets and two lepton subsystem would have been very useful, here we advocate the use of the asymmetric $M_{T2}(b \ell \ell)$, which inherits nearly all the properties of the symmetric one, with an added benefit of larger statistics from tagging just one $b$ jet. By definition, $M_{T2}$ distribution has a kinematic end point that depends on the decay topology. It is observed to have comparable efficiency with that of $M_{b \ell}$.

% = = = = = = = = = = = = = = = = = = = = = = = = = = = = = = = = = = = = = = = = = = = = = = = = = 
%%====================================
  \begin{figure}[t]
%\begin{center}
 \includegraphics[scale=0.6,keepaspectratio=true]{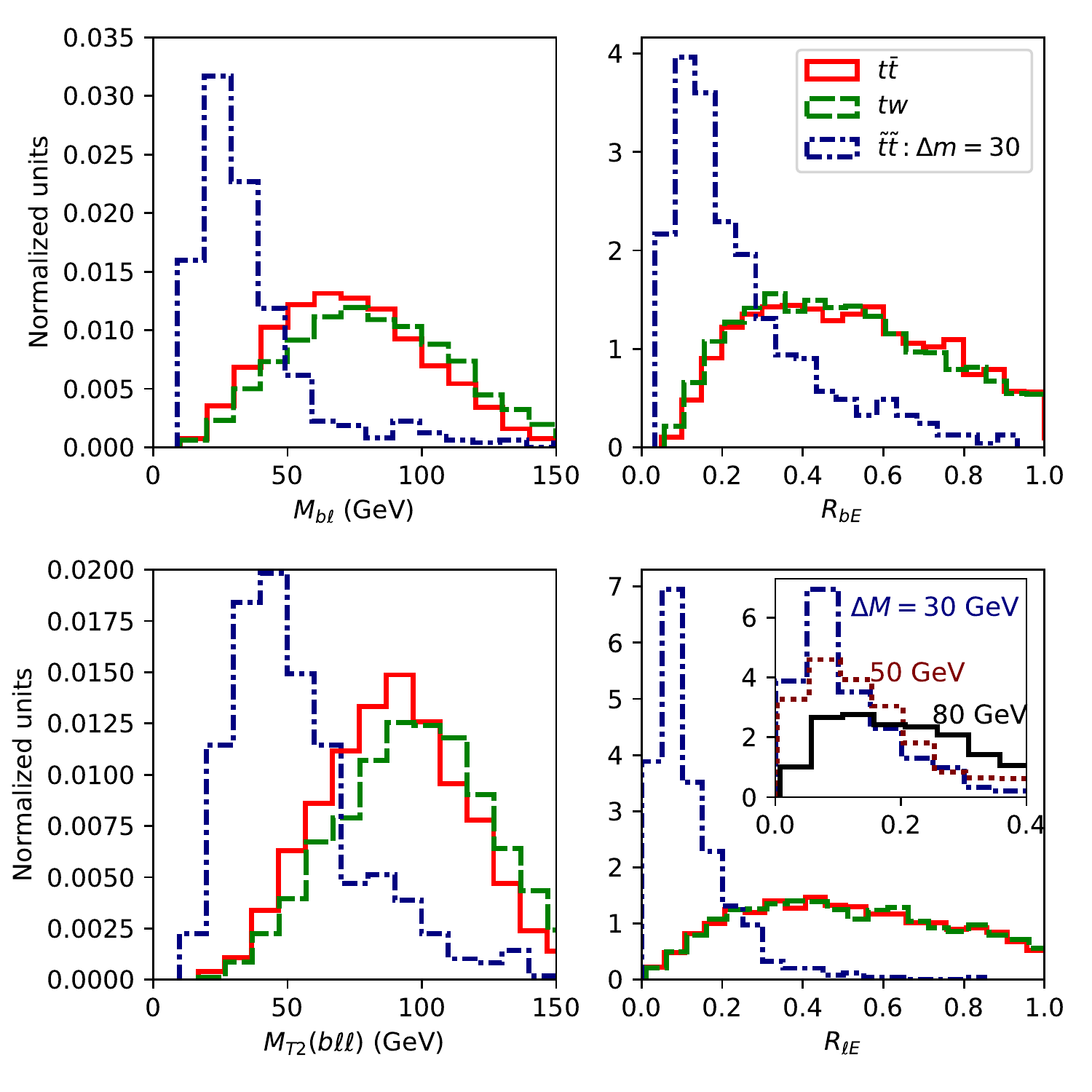}
 % StopFourBodyDecayVars.pdf: 0x0 pixel, 300dpi, 0.00x0.00 cm, bb=
  \vskip -0.4cm
 \caption{The distribution of four variables $M_{b \ell}$, $R_{b E}$,  $R_{\ell E}$, and $M_{T2}(b \ell \ell)$ 
is shown here (clockwise). The blue(dash-dotted), red(solid), and green(dashed) distributions are for the signal, 
$t\bar t$, and $tW$ events, respectively. The distribution was simulated for $\Delta M = 50~ \text{GeV}$ 
and $m_{\tilde{t}} = 400~\text{GeV}$.}
 \label{fig:variables}
%\end{center}
\end{figure}
%%====================================

%%%%%%%%%%%%%%%%%%%%%%%%%%%%%%%% 
\section{Event simulation and basic cuts}
\label{sec:simulation}
%%%%%%%%%%%%%%%%%%%%%%%%%%%%%%%%
%\textit{Event simulation and basic cuts:-} 
We simulate both the signal and the background events using  $\texttt{MadGraph5\_aMC@NLO}$~\cite{Alwall:2014hca}  and those events were passed to \texttt{Pythia8}~\cite{Sjostrand:2006za,Sjostrand:2007gs} for multiparton interaction, hadronization,  and parton showering. Finally, full detector-level simulation is done in \texttt{Delphes3}~\cite{deFavereau:2013fsa} using the CMS card.  Leptons are isolated and should have minimum $p_T$ of 3.5(5) GeV if it is a muon(electron). For such low $p_T$ lepton to make our analysis robust we use the same light leptons selection efficiencies (categorized in the $p_T-\eta$ plane) as reported by the CMS collaboration\cite{light-lepton-efficieny} for the study of Ref.~\cite{CMS-PAS-SUS-16-025}. For $b$ tagging we have used the combined secondary vertex algorithm at the medium operating point(CSVM), which has b-tagging efficiency of approximately 70\%  with light-parton misidentification probability of only 1.5\% \cite{Chatrchyan:2012jua}. All the samples are matched up to  one jet using the MLM scheme\cite{Mangano:2006rw,Hoche:2006ph} where all the jets are reconstructed using anti-kT algorithm with $R = 0.4$ having $p_T > 20$ GeV. The highest $p_T$ non-b jet is tagged as the ISR jet provided  $p_T(j_{ISR}) > 100$ GeV, a modest choice, in comparison to usual compressed searches, to increase the available number of signal events for investigating our variables further. For all our analysis we have used the NLO+NLL top squark cross sections given by the LHC SUSY Cross Section Working Group~\cite{LHC_stop_xsection,Borschensky:2014cia}. The predicted $t\bar{t}$ production cross section is $\sigma_{t\bar{t}}$ = 815.96 pb as calculated with the Top++2.0 program assuming a top quark mass $m_{t}$ = 173.2 GeV~\cite{Czakon:2011xx}. For the $tW$ channel the NLO+NNLL cross section is 71.7 pb\cite{Kidonakis:2015nna}.

Since   most of the $p_T(\tilde{t})$ is carried away by the neutralino, we choose a large $\met$ cut of 250 GeV to reduce a significant amount of background events including the QCD multijet backgrounds. Exploiting the fact that the ISR will be approximately in the opposite direction to the $\met$,  we introduce an additional cut that $|\phi(ISR) -\phi(\met) -\pi| < 0.5$. This will also significantly diminish the enormous QCD background. To minimize the effect of jet mismeasurement contributing to $\met$ we also demand that  $|\phi(j)-\phi(\met)|>0.2$ for all jets other than the ISR.

% = = = = = = = = = = = = = = = = = = = = = = = = = = = = = = = = = = = = = = 
%%%%%%%%%%%%%%%%%%%%%%%%%%%%%%%% 
\section{Results}
\label{sec:results}
%%%%%%%%%%%%%%%%%%%%%%%%%%%%%%%%
%\textit{Results:- }
 \begin{table*}[t]
 \centering
 \begin{tabular}{|c||c|c|c||c|c|}
 \hline
  Cut                                 & \multicolumn{3}{c||}{Signal}                                                                    & \multicolumn{2}{c|}{Background}                                 \\ \hline\hline
                                & $\Delta M = 30$ GeV           & $\Delta M = 50$ GeV           & $\Delta M = 80$ GeV            & $t\,\bar{t}$                   & $t\,W$                         \\ \hline
 Preselection +                     & \multirow{2}{*}{489  [$100\%$]} & \multirow{2}{*}{835  [$100\%$]} & \multirow{2}{*}{1411  [$100\%$]} & \multirow{2}{*}{5165  [$100\%$]} & \multirow{2}{*}{4236  [$100\%$]} \\ 
 $2\ell + ISR+\geq 1\,b$            &                               &                               &                                &                                &                                \\ \hline
 $\met >$ 250 GeV                     & 194  [$39.7\%$]                 & 290  [$34.7\%$]                 & 450  [$31.9\%$]                  & 246  [$4.8\%$]                   & 363  [$8.6\%$]                   \\ \hline
 $|\phi(ISR) -\phi(\met) -\pi| < 0.5$ & 170  [$34.8\%$]                 & 249  [$29.8\%$]                 & 406  [$28.8\%$]                  & 198  [$3.8\%$]                   & 278  [$6.6\%$]                   \\ \hline
 $M_{b\ell} < 60$                       & 134  [$27.4\%$]                 & 216  [$25.9\%$]                 & 319  [$22.6\%$]                  & 64  [$1.2\%$]                    & 52  [$1.2\%$]                    \\ \hline
 $R_{bE} < 0.2$                          & 120  [$24.5\%$]                 & 187  [$22.4\%$]                 & 290  [$20.6\%$]                  & 32  [$0.6\%$]                    & 29  [$0.7\%$]                    \\ \hline
 $R_{\ell E} < 0.3$                      & 120  [$24.5\%$]                 & 185  [$22.2\%$]                 & 272  [$19.3\%$]                  & 22  [$0.4\%$]                    & 13  [$0.3\%$]                    \\ \hline
 \end{tabular}
\caption{Effectiveness of the kinematic variables to minimize the colossal background is represented here. Using these variables it is possible to retain at least 20$\%$ of the signal events while discarding 99.6$\%$ of the background events. The signal events are generated keeping the stop mass at 400 GeV.}
 \label{tab:cutflow}
 \end{table*}

Using the simulated events that passed all the basic selection cuts described above, we have plotted all four pivotal variables for both signal and background events in Fig.~\ref{fig:variables}. While performance of $M_{b\ell},\,M_{T2}(b\ell\ell)$ and $R_{bE}$ in the four-body decay region remains robust, the $R_{\ell E}$ distribution starts spreading for larger $\Delta M$, {\it e.g.}, typically for 50 GeV or more. This is shown in the inset of the bottom-right plot of Fig.~\ref{fig:variables}. Hence, we preach using a different $R_{\ell E}$ cut for different mass gap and we have used the value best suited for $\Delta M = 80$ GeV, reported later in this section. 
As we can see both the kinematic variables $R_{b E}$  and $R_{\ell E}$ fall sharply for the signal events whereas the background $t\bar{t}$ distribution is rather flat. As described before this is due to large $\met$ and very soft leptons/$b$ jets originating from the decay of the top squark. The ratio $R_{bE}$ shows a little higher value because of two reasons: we are taking just one $b$ jet contribution and also there are chances that the origin of the $b$ jet is not from the top squark decay. 
  
In Fig.~\ref{fig:variables} we also plotted two mass variables, namely the invariant mass of $b\ell$ system $M_{b\,\ell}$ and the stransverse mass distribution $M_{T2}$ of the $b \ell \ell$ asymmetric subsystem setting the input trial invisible mass as our trivial choice, 0. Although we have not used the $M_{T2}(b\ell\ell)$ variable for present analysis, it was also tested as a good discriminator  with $M_{T2}(b\ell\ell) < 70 $ GeV being a signal rich region. One can notice that for signals most of the events having smaller invariant mass $M_{b \ell}$ compared to the background makes this a good discriminator. The $M_{b \ell}$ distribution for signal has a tail instead of an end point at $\Delta M$; this is because of the detector resolution and other realistic effects. There are two invariant masses possible using two leptons and one $b$ quark and we take the smaller one among them.

In order to maximize signal to background ratio we optimized the event selection cuts as
(i) $M_{b \ell}  < 60$ GeV, 
(ii) $R_{b E}    < 0.2$,  and
(iii) $R_{\ell E}< 0.3$. 

The new variables are indeed remarkably effective in minimizing the colossal background and this is one of the main results of the study. It is possible to retain at least 20\% of the signal events while discarding 99.6\% of the background events. The relevant cut flow for both signal and background is shown in Table~\ref{tab:cutflow}.

We emphasize that the variables we have proposed are neither unique to the signal we are analyzing here nor dependent on this specific topology. They can be exploited for other search channels too.

%%====================================
\begin{figure}[t]
\includegraphics[angle=0,scale=0.3,keepaspectratio=true]{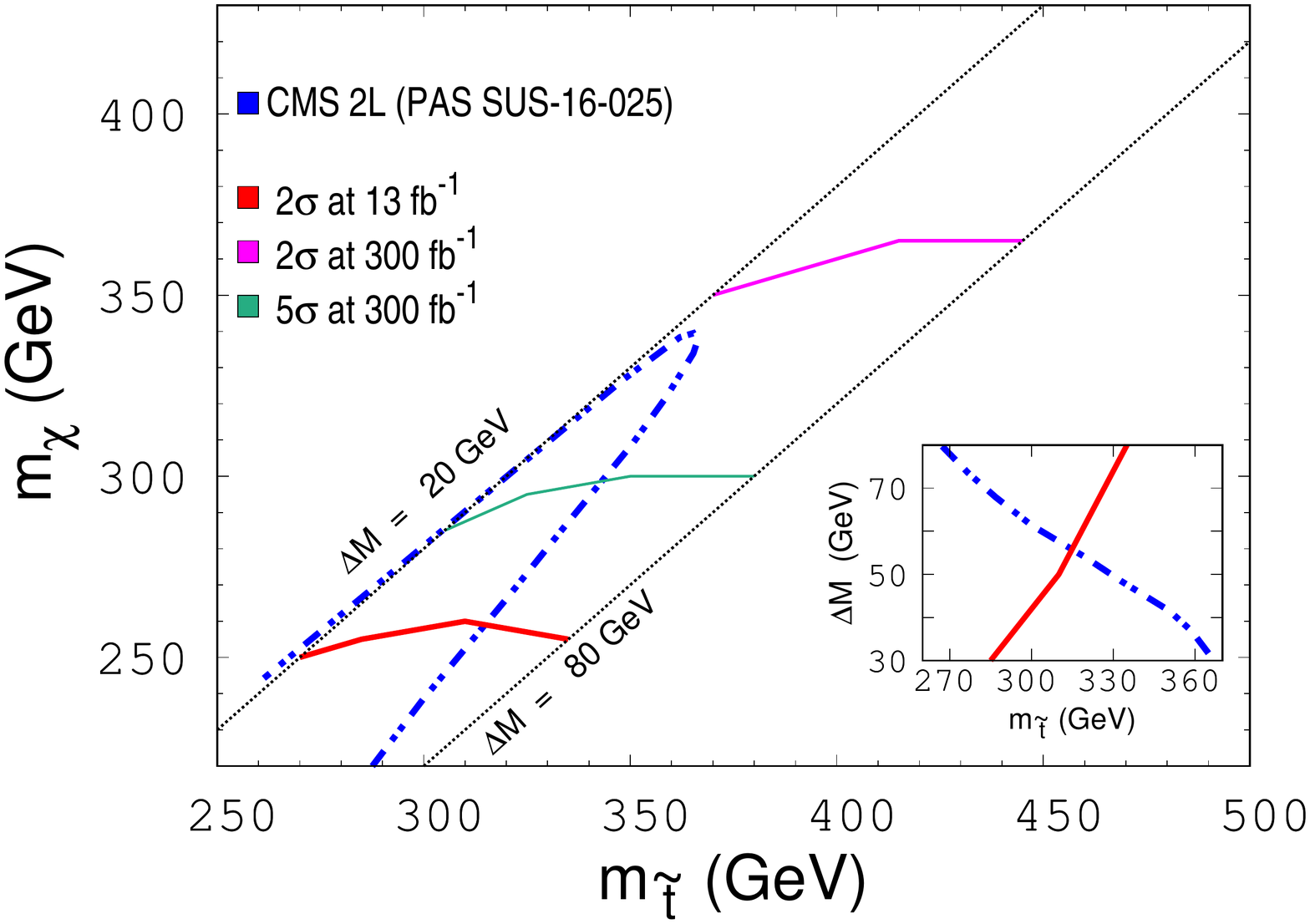} % width=9cm
 \vskip -0.1cm 
\caption{\label{fig:sensitivity}
 Exclusion limit with 13 and 300 $fb^{-1}$ data and discovery plot for 300 $fb^{-1}$ data are shown. 
The inset shows exclusion limit in the $m_{\tilde{t}}-\Delta M$ plane with 13.3 $fb^{-1}$ luminosity.}
\end{figure}
%%====================================

Equipped with the optimized variables and event selection criteria, we analyze the signal as well as background events, and plot the statistical significance for 13 TeV LHC with $13\textrm{ and }300\,fb^{-1}$ data in Fig.~\ref{fig:sensitivity}. For 2$\sigma$ significance we use the formula $\sqrt{2\left[(S+B)\textrm{ln}\left(1+\frac{S}{B}\right)-S\right]}$ and for $5\sigma$ the corresponding formula is ${S}/{\sqrt{B}}$ where $S(B)$ stands for number of signal(background) events at a particular integrated luminosity. The blue (dash-dot) curve shows the state-of-the-art limit on the $\tilde{t}$ mass coming from the dilepton search as presented by the CMS collaboration~\cite{CMS-PAS-SUS-16-025}.  As we can see the limit is poor for larger mass difference and drops down pretty fast as we move towards the $\Delta M = m_W $ boundary. The red solid curve shows our exclusion limit at 13 TeV LHC with 13$fb^{-1}$ data. Evidently the new kinematic and mass variables work rather well for this compressed parameter space. In particular, in the higher mass gap side our search channel provides a better limit and can act as an excellent compliment search to the existing CMS search. In the inset of Fig.~\ref{fig:sensitivity} we have shown the results for luminosity of 13$fb^{-1}$ in the $m_{\tilde{t}}-\Delta M$ plane. It is evident from the inset that the larger mass gap region can effectively be probed using the new variables proposed here. 

Also, we exhibit the limits for 13 TeV LHC with an integrated  luminosity of 300 $fb^{-1}$. The magenta curve shows 2$\sigma$ exclusion limits whereas the green curve shows 5$\sigma$ discovery potential. With our proposed variables at 13 TeV with 13 $fb^{-1}$ data we can exclude the top squark up to 335 GeV with neutralino mass 255 GeV and with integrated luminosity of 300 $fb^{-1}$ data the limit on the top squark can go up to 445 GeV with $m_\chi$ = 365 GeV. Also, it is possible to discover the much sought top squark with 300 $fb^{-1}$ data if the top squark lies below 380 GeV with a mass gap of 80 GeV.  

Regardless of the smaller branching ratio in the leptonic channel, proposed variables are capable of delivering limits comparable to that of hadronic modes~\cite{An:2015uwa}. 
% using 300 $fb^{-1}$ data.} %  which are very much
%
In fact, these variables are  not limited to the four-body decay only. One can exploit them for studying other possible decay modes in compressed SUSY~\cite{Konar:2017oah}.

% = = = = = = = = = = = = = = = = = = = = = = = = = = = = = = = = = = = = = = 
%%%%%%%%%%%%%%%%%%%%%%%%%%%%%%%% 
\section{Conclusion and discussion}
\label{sec:conclusion}
%%%%%%%%%%%%%%%%%%%%%%%%%%%%%%%%
%\textit{Conclusion and discussion:- } 
In this article, we consider the near degenerate top squark with the lightest neutralino where the top squark undergoes four-body decay. Among all decay channels the dileptonic mode, despite being an experimentally cleaner and more reliable search channel, is challenging because of two additional neutrinos present in the final state and also due to low branching ratio. The present experimental limit is rather weak compared to the hadronic search despite the former having a clear advantage of identifying the isolated leptons along with the $b$ quark. We proposed suitable kinematic variables that best exploit the decay topology information producing the kinematic end points. These observables include invariant mass, stransverse mass, and two new ratios that discriminate the signal from the background efficiently.

With these variables the existing limit on the top squark can be extended up to 335 GeV for mass gap of 80 GeV with  integrated luminosity of $13 fb^{-1}$ at $95\%$ confidence level.  Evidently, this approach provides a complementary search strategy to the present CMS analysis for higher mass gap region where the current exclusion limit is 270 GeV. Hence, we advocate to implement the proposed variables to enhance the observables capability of the LHC experimental collaborations.

% = = = = = = = = = = = = = = = = = = = = = = = = = = = = = = = = = = = = = = = = = = = = = = = = = 
\bigskip
\acknowledgments
%  {\emph{Acknowledgments}.}---
%\begin{acknowledgments}
We thank A.  Bhardwaj for validating our work using MadAnalysis5. This work was supported by the Physical Research Laboratory (PRL), Department of Space (DOS), India. T.M. acknowledge the funding available from the Department of Atomic Energy, Government of India, for the Regional Centre for Accelerator based Particle Physics (RECAPP), Harish-Chandra Research Institute.
%\end{acknowledgments}

%\appendix
% = = = = = = = = = = = = = = = = = = = = = = = = = = = = = =
%\section{Cut efficiency and usefulness of $R$ variables}
%\label{app:suppl}

% = = = = = = = = = = = = = = = = = = = = = = = = = = = = = 

%%%%%%%%%%%%%%%%%%%%%%%%%%% Bibliography %%%%%%%%%%%%%%%%%%%%%%%%%%%%%%%
% \bibliographystyle{unsrt}
\bibliographystyle{apsrev4-1}
\bibliography{bibliography}

\end{document}